\documentclass[12pt, manuscript]{aastex}
\usepackage{natbib}
\usepackage{color}
\usepackage{comment}

\begin{document}
\title{Data Constrained Coronal Mass Ejections in A Global Magnetohydrodynamics Model}
\author{M. Jin\altaffilmark{1,2}, W. B. Manchester\altaffilmark{3}, B. van der Holst\altaffilmark{3}, I. Sokolov\altaffilmark{3},  G. T\'{o}th\altaffilmark{3}, R. E. Mullinix\altaffilmark{4}, A. Taktakishvili\altaffilmark{4,5}, A. Chulaki\altaffilmark{4}, and T. I. Gombosi\altaffilmark{3}}

\altaffiltext{1}{Lockheed Martin Solar and Astrophysics Lab, Palo Alto, CA 94304, USA; jinmeng@lmsal.com}
\altaffiltext{2}{NASA/UCAR LWS Jack Eddy Fellow}

\altaffiltext{3}{Climate and Space Sciences and Engineering, University of Michigan, Ann Arbor, MI 48109, USA; chipm@umich.edu}

\altaffiltext{4}{Community Coordinated Modeling Center, NASA Goddard Space Flight Center, Greenbelt, MD 20771, USA; richard.e.mullinix@nasa.gov}

\altaffiltext{5}{Catholic University of America, Washington, DC 20064, USA; Aleksandre.Taktakishvili-1@nasa.gov}

\begin{abstract}
We present a first-principles-based coronal mass ejection (CME) model suitable for both scientific and operational purposes by combining a global magnetohydrodynamics (MHD) solar wind model with a flux rope-driven CME model. Realistic CME events are simulated self-consistently with high fidelity and forecasting capability by constraining initial flux rope parameters with observational data from GONG, SOHO/LASCO, and STEREO/COR. We automate this process so that minimum manual intervention is required in specifying the CME initial state. With the newly developed data-driven Eruptive Event Generator Gibson-Low (EEGGL), we present a method to derive Gibson-Low (GL) flux rope parameters through a handful of observational quantities so that the modeled CMEs can propagate with the desired CME speeds near the Sun. A test result with CMEs launched with different Carrington rotation magnetograms are shown. Our study shows a promising result for using the first-principles-based MHD global model as a forecasting tool, which is capable of predicting the CME direction of propagation, arrival time, and ICME magnetic field at 1 AU (see companion paper by \citealt{jin16b}).
\end{abstract}


\section{introduction}
Coronal mass ejections (CMEs) are a major source of potentially destructive space weather conditions (e.g., geomagnetic storms, solar energetic particles). Due to our increasing dependence on advanced technology, which is vulnerable to severe space weather conditions, there is a high national priority to establish a reliable space weather forecasting capability. However, available CME observations that may provide a basis for forecasts are very limited.  In particular, erupting magnetic fields cannot be directly observed in the solar corona, which is of critical importance given that interplanetary magnetic field (IMF) is a major driver of the geomagnetic storms. For these reasons, there is a great need to be able to predict CME magnetic fields based on photospheric observations, both for scientific understanding and for forecasting purposes.

In the past two decades, many CME forecasting models have been developed, which can be divided mainly into three different categories. The first category is empirical forecasting models, which use near-Sun CME observations to estimate the arrival time of CMEs at 1 AU through empirical relations built through a large number of observations (e.g., \citealt{gopalswamy01}). Recently, the data-mining techniques have been utilized to establish empirical models (Riley et al. 2015). The second category is called kinematic models, which solve a reduced form of fluid equations without dynamics. Successful examples include the 3-D Hakamada-Akasofu-Fry version 2 (HAFv.2) model, in which remote-sensing data is used to derive the shock speed and direction \citep{hakamada82, fry01, dryer04}, and the cone model, which fits CME observations with three free parameters: angular width, speed, and central CME position \citep{zhao02, hayashi06}. The cone model has been widely used by the research community to predict the CME/CME-driven shock velocity (e.g., \citealt{xie04, michalek07, luhmann10, vrsnak14}). In order to provide more accurate forecasts as well as the CME plasma parameters (i.e., density, temperature, velocity) in addition to the arrival time, the third category of forecasting models couple kinematic models with heliospheric magnetohydrodynamics (MHD) models (inner boundary outside magneto sonic point). In this case, the kinematic models are used to prescribe the initial conditions of the MHD models. By combing the cone model with Enlil \citep{odstrcil05}, the average error in the CME Analysis Tool (CAT; \citealt{millward13})-Wang-Sheeley-Arge (WSA; \citealt{arge00})-Enlil operational forecasting model at Space Weather Prediction Center (SWPC) is $\sim$7.5 hours, which represents the state-of-the-art CME forecasting model in operation. 

More recent MHD models introduce the flux rope beyond the Alfv\'{e}n surface allowing for prediction of the magnetic field at 1 AU. By utilizing a spheromak CME model in an MHD inner heliosphere solar wind model, \citet{shiota16} demonstrated that the model is capable of predicting magnetic profile at the Earth. However, since the model starts outside the Alfv\'{e}n surface, the CME model is less constrained by observations. To provide magnetic field forecasting, several models have been developed which self-consistently simulate the CME from the corona to 1 AU by combining realistic MHD corona models (e.g., \citealt{mikic99, groth00, rou03, cohen07, feng11, evans12, sokolov13, bart14}) and magnetically driven eruptions (e.g., \citealt{gibson98, titov99, anti99, titov14}). This approach represents the most sophisticated CME propagation model so far. And many case studies have provided high fidelity simulations both at the Sun and at 1 AU (e.g., \citealt{chip04b, lugaz07, toth07, cohen08, lionello13, chip14, shen14, jin16b}).

As the first step to transfer a research model to operational forecasting, it is important to characterize the behavior of the magnetic-driven eruptions in the ambient solar wind solutions. Using WSA-Enlil modeling system, \citet{pizzo15} suggest that the CME propagation is non-chaotic and relates to a finite set of inputs in the MHD simulations. Therefore, it is possible to control the CME behavior through varying the initial CME parameters to match observations. In our study, we follow this path by combing the Alfv\'{e}n Wave Solar Model (AWSoM; \citealt{sokolov13, oran13, bart14}) with the Gibson-Low (GL) analytical flux rope model, and we present a method to determine the initial flux rope parameters from available observations so that the CME speed can be well reproduced near the Sun. With the CME structure self-consistently propagating into the heliosphere, the model has the capability to follow the erupting magnetic field from the Sun to 1 AU and provide a forecast of CME arrival time and all plasma parameters (i.e., velocity, density, temperature, and magnetic field).

The paper is organized as follows: In Section 2, we briefly describe the AWSoM model used to construct the background solar wind and the GL flux rope model for the CME initiation. The methodology of determining the GL flux rope parameters will be described in detail in Section 3, followed by the discussion and summary in Section 4.

\section{Models}
Here, we describe the Eruptive Event Generator using Gibson-Low configuration (EEGGL), an automated tool for finding parameters of the GL flux rope to reproduce observed CME events. In Figure 2, an example of the GL flux rope is shown embedded into the global solar wind solution of Carrington Rotation (CR) 2107. Figure 4 shows a chart demonstrating how CME events can be simulated in the AWSoM. First, the synoptic magnetogram is used to specify the inner boundary condition of the magnetic field for AWSoM, which is then employed to generate a steady state solar wind solution. At the same time, the input magnetogram and observed CME speed (from SOHO/LASCO and/or STEREO/COR observations) are used by EEGGL to determine the GL flux rope parameters. With the derived parameters, a GL flux rope is inserted into the steady state solar wind to initiate the CME event. In this Section, we briefly describe the AWSoM and GL analytical flux rope model.

\subsection{Background Solar Wind Model}
The MHD solar wind model used in this study is the recently developed AWSoM \citep{sokolov13, oran13, bart14}, which is a data-driven model with a domain extending from the upper chromosphere to the corona and heliosphere. A steady-state solar wind solution is obtained with local time stepping and second-order shock-capturing scheme \citep{toth12}. The inner boundary condition for the magnetic field can be specified by different magnetic maps (e.g., from GONG, SOHO/MDI, or SDO/HMI). In this study, the magnetograms from GONG are used. The inner boundary conditions for electron and proton temperatures $T_{e}$ and $T_{i}$ and number density $n$ are assumed at  $T_{e}=T_{i}=$ 50,000 K and $n =$ 2$\times$10$^{17}$ m$^{-3}$, respectively. At the base of the atmosphere, the temperature is held fixed at 50,000 K while the density falls off exponentially until it reaches a level where the radiative losses are sufficiently low that the temperature increase monotonically with height.  Above this height, the temperature increases rapidly forming the transition region.  This procedure allows chromospheric evaporation to self-consistently populate the corona with an appropriately high plasma density. The inner boundary density and temperature do not otherwise have a significant influence on the global solution \citep{lio09}. The initial conditions for the solar wind plasma are specified by the Parker solution \citep{parker58}, while the initial magnetic field is based on the Potential Field Source Surface (PFSS) model solved with the Finite Difference Iterative Potential Solver (FDIPS, \citealt{toth11}).

In the AWSoM, Alfv\'{e}n waves are represented by energy densities of counter-propagating waves. The dissipation is based on the interaction between the counter-propagating waves with self-consistent wave reflection. Alfv\'{e}n waves are specified at the inner boundary with a Poynting flux that scales with the surface magnetic field strength. The solar wind is heated by Alfv\'{e}n wave dissipation and accelerated by thermal and Alfv\'{e}n wave pressure. Electron heat conduction (both collisional and collisionless) and radiative cooling are also included in the model. These energy transport terms are important for self-consistently creating the solar transition region. In order to produce physically correct solar wind and CME structures, such as shocks, the electron and proton temperatures are treated separately \citep{chip12, jin13}. Thus, while the electrons and protons are assumed to have the same bulk velocity, heat conduction is applied only to the electrons, owing to their much higher thermal velocity \citep{kos91}. 

By using a physically consistent treatment of wave reflection, dissipation, and heat partitioning between electrons and protons, the AWSoM has demonstrated the capability to reproduce the solar corona environment with only 3 free parameters that determine Poynting flux ($S_A / B$), wave dissipation length ($L_\perp\sqrt{B}$), and stochastic heating parameter ($h_S$) \citep{bart14}. In Figure 1, an example of a steady state solar wind solution is shown. Panel (a) shows the steady state solar wind speed of a meridional slice at $X=0$. Panel (b) shows the 3D field configuration near the Sun with large-scale helmet streamer belt and open/active region field lines marked by white and green, respectively. Carrtington coordinates are used, in which the Z axis corresponds with the rotation axis. Also, in Figure 2(b)-(d), a steady state solar wind solution of velocity, proton, and electron temperatures near the Sun are shown. The electron and proton temperatures are very close near the Sun due to collisions and they diverge further away from the Sun. 

\subsection{Gibson-Low Flux Rope Model}
In this study, we initiate CMEs using the analytical Gibson-Low (GL; \citealt{gibson98}) flux rope model implemented in the Eruptive Event (EE) of Space Weather Modeling Framework (SWMF; \citealt{toth12}). This flux rope model has been successfully used in numerous modeling studies of CMEs (e.g., \citealt{chip04a, chip04b, lugaz05a, lugaz05b, schmidt10, chip14, jin16}). The GL flux rope are obtained by finding an analytical solution to the magnetohydrostatic equation $(\nabla\times{\bf B})\times{\bf B}-\nabla p-\rho {\bf g}=0$ and the solenoidal condition $\nabla\cdot{\bf B}=0$. This solution is derived by applying a mathematical stretching transformation $r\rightarrow r-a$ to an axisymmetric solution describing a twisted toroidal flux rope contained within a sphere of radius $r_0$ centered relative to the heliospheric coordinate system at $r=r_1$ (black contour in Figure 2(a)). The full 3D field of $\bf B_{\rm GL}$ can be expressed by a scalar function $A$ and a free parameter $a_1$ that determines the magnetic field strength \citep{lites95}. The transformed flux rope appears as a tear-drop shape of twisted magnetic flux (red contour in Figure 2(a)). At the same time, Lorentz forces are introduced, which leads to a density-depleted cavity in the upper portion and a dense core approximating a filament at the lower portion of the flux rope. This flux rope structure helps to reproduce the commonly observed 3-part density structure of the CME \citep{llling85}. 

In sum, the GL flux rope is mainly controlled by five parameters: the stretching parameter $a$ determines the shape of the flux rope, the distance of torus center from the center of the Sun $r_{1}$ determines the initial position of the flux rope before it is stretched, the radius of the flux rope torus $r_{0}$ determines the size of the flux rope, the flux rope field strength parameter $a_1$ determines the magnetic strength of the flux rope, and a helicity parameter to determine the positive (dextral)/negative (sinistral) helicity of the flux rope. The relative magnitudes of B$_\phi$ and B$_\theta$ at any point in the flux rope model are of fixed functional forms such that the magnetic field makes a transition from pure toroidal field at the core to pure poloidal field at the outer surface.

The GL flux rope and contained plasma are then superposed onto the steady state solar corona solution: i.e. $\rho=\rho_{0}+\rho_{\rm GL}$, ${\bf B = B_{0}+B_{\rm GL}}$, $p=p_{0}+p_{\rm GL}$. The combined background-flux rope system is in a state of force imbalance (due to the insufficient background plasma pressure to offset the magnetic pressure of the flux rope), and thus erupts immediately when the numerical model is advanced forward in time. In Figure 2(a), the GL flux rope in 2D is shown with magnetic field lines and plasma density. We can see that the main flux rope structure is embedded in a coronal streamer with high plasma density while a low density cavity fills the outer part of the rope and dense filament material fills the bottom. In Figure 2(b)-(d), the 3D GL flux rope is shown with velocity, proton temperature, and electron temperature on the central meridional plane of the flux rope. 

In Figure 3, we show three GL flux ropes with different size ($r_0$) and magnetic field strength ($a_1$) parameters. The left panel shows the initial configuration of the GL flux ropes. The blue and red isosurfaces represent a density ratio of 0.3 and 2.5 between the solutions before and after the flux rope insertion. The middle panel shows the resulting CME evolution at 10(20) min. The background color shows the density ratio between the solution at 10(20) min and the background solar wind. The red, white, and green field lines represent CME flux rope field lines, large-scale helmet streamers, and field lines from surrounding active regions and open field. The right panel shows the synthesized SOHO/LASCO white light images. The color scale shows the white light total brightness divided by that of the pre-event background solar wind. The first ($r_0=0.8$, $a_1=0.6$) and second ($r_0=0.8$, $a_1=2.25$) cases have the same flux rope size but different magnetic field strength. Comparing Figure 3(a) and 3(d), we can see that with a higher magnetic field strength parameter, more plasma is added at the bottom of the flux rope (red isosurface). The slight size difference is due to the background field. Since the GL flux rope solution is added to the background solar wind solution, the field near the GL flux rope boundary may be affected with weak magnetic field strength parameter. After initiation, the higher flux rope magnetic field leads to a higher CME speed and stronger density pile-up in front of the flux rope. The second and third ($r_0=0.6$, $a_1=2.25$) cases have the same magnetic field strength parameter but with different flux rope size. In this case, we can see the flux rope is considerably smaller at the beginning. With this smaller flux rope, the resulting CME speed is reduced and the morphology of CME in the synthesized white light image is quite different with narrower CME width angle.

There are several advantages of using this force imbalanced flux rope for this study. First, the flux rope fits into an active region with free energy for eruption, which does not require time-consuming energy build-up process in the simulation. Second, the GL flux rope comes with dense plasma over the polarity inversion line (PIL) and a low density cavity above it, which mimics CME observations. Third, the eruption speed can be controlled by the GL flux rope parameters. Moreover, previous studies have shown that the GL flux rope is capable of producing magnetic cloud signature at 1 AU in the simulations (e.g., \citealt{chip04b, chip14}).

\section{Constraining Flux Rope Parameters with Observations}
EEGGL is designed to determine the GL flux rope parameters, including flux rope location, orientation, and five parameters to control the characteristics of the flux rope in order to model specific CME events. EEGGL operates under the assumption that the CMEs originate from closed corona flux forming loop structures anchored to the photosphere in bipolar active regions. From EEGGL's graphical interface (shown in Figure 5), the user selects the active region from which the CME originates. EEGGL then calculates the weighted center for both positive and negative polarities of the source region. At the same time, EEGGL determines the PIL locations of the source region. The flux rope location then can be determined by the intersection between the PIL and the line connecting positive and negative weighted centers. Also, the orientation of the flux rope is set to be parallel to the PILs. In the cases that the PILs are complex curves, a straight line is fitted using the PIL points near the flux rope center to represent the PILs.  In case of non-bipolar regions, the PIL that is responsible for the eruption will be used to determine the orientation of the flux rope. One example of non-bipolar source region is the AR0 in Figure 5. The PIL of AR0 has an up-side-down ``L" shape. When the eruption site is determined from the observation, the GL flux rope orientation and the source region size are calculated by the erupting part of the PIL. The flux rope helicity (dextral/sinistral) is determined according to the hemispheric helicity rule that the dextral/sinistral filament dominates the northern/southern hemisphere. A recent study by \citet{liu14} found that $75\%\pm7\%$ of 151 ARs observed with HMI obey this helicity rule. Several examples are shown in Figure 5, in which the GONG Carrington magnetogram of CR 2107 is shown\footnote{The resolution of the magnetogram is 360$\times$180. The latitude grid is evenly spaced in sine latitude.}. The active region markers (AR0 -- AR6) represent the CME source regions we used for the parameter study. Note that, to reduce the complexity of the photospheric magnetic field configuration, we smooth the magnetogram by a 5$\times$5 pixel window.  

Of the four GL flux rope parameters, EEGGL calculates the flux rope size $r_{0}$ and flux rope field strength $a_{1}$ and fixes the other two parameters ($r_{1}$=1.8, $a$=0.6). The reason for fixing these two parameters is that they cannot be easily related to or obtained from normal observations. On the other hand, by varying these parameters, two main CME characteristics (CME width and speed) can be adjusted to match observations. The fixed values were selected a priori for optimal performance, in particular how it relates to the amount of disconnected flux and the CME mass. The mass falls in the range of typical CMEs, of the order of 10$^{15}$ grams \citep{chip04b} and increases with field strength. This is the best we can do, since we can not regularly determine the filament mass of the CME. In order to find a useful empirical relationship between the GL flux rope parameters and the resulting CME speeds, we examined the observational data in \citet{qiu07} where we found an approximately linear relationship between CME speed and reconnected flux. The data from \citet{qiu07} is replotted in Figure 6. From this result, we searched for an empirical relationship between CME speed and poloidal magnetic flux of the GL flux rope. The poloidal flux can be calculated by integrating $B_\phi$ over a central surface across magnetic axis of GL flux rope. The expression of $B_\phi$ can be found in the Appendix of \citet{gibson98}. With the two GL flux rope parameters fixed ($a$ and $r_1$), the GL poloidal flux is only determined by the size ($r_0$) and magnetic strength ($a_1$) parameters: $\Phi_{Poloidal}=c\cdot a_{1} r_{0}^{4}$, where $c$ is a constant.

We initiate CMEs with different flux rope sizes and magnetic field strengths and run each simulation 30 minutes after the flux rope insertion. We then derive the CME speeds in the simulation taken as the average speeds of the CME at the outer-most front between 20 and 30 minutes. The outermost front of the CME is determined by finding the CME propagation plane and extracting the line profiles along the CME propagation path (see \citealt{jin13} for an example). The typical plasma parameters we use to identify the outermost front are proton temperature, velocity, or density profiles. The 20-30 min window is chosen because during that period of time the CME is reaching a nearly constant speed both in the simulation and in the observation. Therefore, the speed obtained is more reliable and consistent. According to the method we used to determine the CME speed, the major source of uncertainty comes from the grid size at the outermost front in the simulation that can affect the determination of the exact outermost point on the profile (e.g., Figure 5 and 6 in \citealt{jin13}). Therefore, we estimate the uncertainty in velocity by $(dr_1+dr_2)/2 \cdot dt$, where $dr_1$ and $dr_2$ are the CME front grid size at 20 min and 30 min, respectively. $dt$ is 600 s in this case.

To investigate the relationship between the poloidal flux of the GL flux rope and the resulting CME speed, we initiate 8 CMEs from AR0 (marked in Figure 5) with different flux rope size and strength parameters. The CME parameters can be found in Table 1. The poloidal fluxes vary from $6.16\times10^{21}$ to $2.0\times10^{22}$ Mx, which lead to CME speeds varying from 758 km s$^{-1}$ to 3150 km s$^{-1}$. In Figure 6, the relationship between the input poloidal fluxes and the resulting CME speeds in the simulation is shown, in which we find a very good linear relationship between the two variables. The red dashed line shows a linear fit result. 

We next investigate how the same flux rope behaves in different active regions. We test all the 7 major active regions in CR2107 magnetogram (marked in Figure 5) and the results are shown in Table 2 and Figure 7(b)/(d). The flux rope parameters ($r_{0}$=0.8, $a_{1}$=1.6) are fixed for all the 7 runs. The average magnetic field $B_{r}$ (measured both around the flux rope center and along the PIL of the source region) is used to quantify the source region. With the same flux rope, we find that the CME speed is inversely related to the average $B_{r}$ of the active region through a nonlinear relationship that is not as well defined as the CME speed-poloidal flux relationship (Figure 6). The spread of points suggest a dependence of CME speed on more complex features of the active region that are not captured by the average field strength.

With these two empirical relationships obtained from the parameter study, it is possible to derive an equation to calculate the GL flux rope parameters based on the observed CME speed and average magnetic field of the source region:
\begin{equation}
a_{1}=\frac{v_{cme}\cdot \overline{B_r}^\alpha-\beta}{\gamma\cdot r_{0}^4}
\end{equation}
where $a_{1}$ is the parameter that determines the GL flux rope field strength, $v_{cme}$ is the observed CME speed near the Sun, $\overline{B_r}$ is the area average radial field strength of the source region ($\overline{B_r}=\frac{\int B_{r} dA}{A}$), $r_{0}$ is the size of the GL flux rope, $\alpha$, $\beta$ and $\gamma$ are the fitting constants. The observational CME velocity can be obtained through StereoCat CME analysis tool\footnote{http://ccmc.gsfc.nasa.gov/analysis/stereo/} at CCMC. This tool provides the measured CME speed through stereoscopic reconstruction using both SOHO/LASCO and STEREO/COR observations. For determining the GL flux rope size, we investigate two different algorithms: (1) relate the flux rope size to the length of the PIL; (2) relate the flux rope size to the size of the source region. Since the complexity of the PILs, the first algorithm is not robust and may lead to extremely large flux ropes. Therefore, we use the second algorithm to calculate the GL flux rope size in this study. In Figure 5, we show all the 7 source region boundaries in cyan contours calculated by the algorithm. In case of very large or very small complex non-bipolar source regions, the GL flux rope size may be over- or under-estimated. We set up a lower and upper limit of GL flux rope size ($r_0\in [0.2,2.0]$). Note that, since the CME speed in the simulation is determined by GL poloidal flux, the uncertainty of flux rope size will not influence the final CME speed.

In Figure 7, we show the fitting results based on the current tests. In Figure 7(a)-(b), the average $B_r$ is calculated around the center of the source region ($\pm 2^{\circ}$ in latitude and longitude). The fitted constants are: $\alpha=0.55\pm0.05$, $\beta=4.13\pm1.05\times10^3$, and $\gamma=1.97\pm0.37\times10^4$. In Figure 7(c)-(d), the average $B_r$ is calculated along the PIL\footnote{The area for calculating average $B_r$ is chosen by $\pm1$ pixel around the PIL.}. The fitted constants are: $\alpha=0.75\pm0.07$, $\beta=3.86\pm0.99\times10^3$, and $\gamma=1.76\pm0.34\times10^4$. The fitting curves based on Eq. (1) are shown in red dashed lines. The fitting results are very close for both methods of determining the average $B_r$. Although the current empirical relationship is obtained from limited number (14) of runs, there is no sudden change in the curves shown in Figure 6 and 7 that needs to be resolved. Therefore, the empirical relationship is expected to be valid for the parameter space of moderate to fast CMEs. We note that the fitted empirical relationship may not work for slow CMEs in a weak source region, which may lead to negative calculated flux rope magnetic strengths. For example, if the source region has an average $B_{r}$ of 10 Gauss, then the input CME speed cannot be less than 740 km s$^{-1}$. The EEGGL will give an error message when negative magnetic strength are calculated.

To further validate this empirical relationship obtained from CR 2107 active regions, we choose 4 different CRs and initiate 5 CMEs using EEGGL calculated flux rope parameters. The input CME speeds into the EEGGL are shown in the third column of Table 3. With each CME simulation runs to 30 min, we calculate the actual CME speed in the simulation taken as the average CME speed at the outer-most front between 20 and 30 minutes, the same definition as CR 2107 runs. Then the simulated and input CME speeds can be compared. The validation result is shown in Table 3, which confirms that the empirical relationship works well for other CRs and different source region configurations. The error (defined by the relative difference between the input and simulated CME speeds) varies from $-4.3\pm14.6\%$ to $+16\pm12.6\%$.

\section{Discussion}
In summary, we present the first data-driven MHD CME model in which the driving flux rope is initiated in the corona and the parameters of the model are constrained by observations near the Sun. We present a new tool, EEGGL, which automatically calculates the parameters to model CME events using synoptic magnetogram data along with the observed CME speed. This CME model and EEGGL are now available at Community Coordinated Modeling Center (CCMC) for runs on demand, where it provides a new forecast capability for the ICME magnetic field at 1 AU that is critically important for space weather. In a companion paper by \citet{jin16b}, we simulate a CME event on 2011 March 7 from the Sun to 1 AU using AWSoM+EEGGL and present detailed analysis of the results to show the new model can reproduce many of the observed features near the Sun and in the heliosphere.
The availability of modeling magnetic eruptions from the low corona to 1 AU also will greatly facilitate research on the coronal and heliospheric responses to CMEs, as well as provide valuable information for improving flux rope driven numerical models including the empirical relationships applied in EEGGL.

For simplicity, the GL flux ropes used in this study have a fixed positive helicity. In future studies, we will set the flux rope helicity based on the vector magnetic field observations and investigate how the helicity of the flux rope affects CME propagation. Active region helicity is available as a data product of the Space-weather HMI Active Region Patches (SHARPs; \citealt{bobra14}) derived from SDO/HMI can now provide the helicity information calculated from the vector magnetic field for all major active regions on the Sun. Therefore, it is possible to obtain the helicity information derived from SHARPs and use it in EEGGL to specify the direction of the toroidal field in the GL flux rope. Moreover, the HEK output of the SDO automatic filament detection \citep{martens12} will be utilized to improve the flux rope size calculation. With increasing number of events studied, the empirical relationship will be improved consistently. Also, more factors (e.g., background density, large-scale magnetic structure) may be considered that affect the CME propagation from the Sun to 1 AU. With more events studied, we could introduce new parameters to account for these effects and further revise the empirical relationship to determine the GL flux rope parameters.

To achieve forecasting, the model must run faster than real time. The computational expense using the current AWSoM model to simulate a CME event from the Sun to 1 AU is about 40,000 CPU hours on the NASA Pleiades supercomputer. This means 1000-2000 CPUs are needed to run the model faster than the real time. With increasing computational power nowadays, the computational requirements may be much more readily available. Moreover, a new model in development, AWSoM-R (Threaded Field Line Model; Sokolov et al. 2016) may run 10-100 times faster than the current AWSoM so that far fewer CPUs will be needed for real time simulations. A preliminary test at the CCMC shows that with 120 CPUs on CCMC cluster hilo, the AWSoM-R steady state simulation takes 17 hours, and the CME evolution of 3 days takes 16 hours.

\begin{acknowledgements}
We are grateful to the anonymous referee for constructive comments that helped improve the paper. M.Jin is supported by NASA/UCAR Jack Eddy Postdoctoral Fellowship and NASA's SDO/AIA contract (NNG04EA00C) to LMSAL. The work performed at the University of Michigan was partially supported by National Science Foundation grants AGS-1322543, AGS1408789 and PHY-1513379, NASA grant NNX13AG25G, the European Union's Horizon 2020 research and innovation program under grant agreement No 637302 PROGRESS. We would also like to acknowledge high-performance computing support from: (1) Yellowstone (ark:/85065/d7wd3xhc) provided by NCAR's Computational and Information Systems Laboratory, sponsored by the National Science Foundation, and (2) Pleiades operated by NASA's Advanced Supercomputing Division. 

This work utilizes data obtained by the Global Oscillation Network Group (GONG) Program, managed by the National Solar Observatory, which is operated by AURA, Inc. under a cooperative agreement with the National Science Foundation. 
\end{acknowledgements}

\newpage
%\bibliographystyle{apj}
%\bibliography{ref}

\newpage
\begin{table}
\caption{\label{tab1}CME Parameter Test}
\begin{center}
\begin{tabular}{ccccc}
\hline \hline Run No. & FR Radius $r_0$ [Rs]  & FR Strength $a_1$\tablenotemark{a} & Poloidal Flux [Mx] & CME speed [km s$^{-1}$]\\
\hline
1  & 0.4 & 2.25 & 1.02$\times$10$^{21}$ & 758$\pm$122 \\
2  & 0.8 & 0.6   & 5.32$\times$10$^{21}$ & 1260$\pm$145 \\
3  & 0.6 & 2.25 &  6.16$\times$10$^{21}$ & 1376$\pm$151 \\
4  & 0.8 & 1.0	& 8.87$\times$10$^{21}$ & 1680$\pm$182 \\
5  & 0.8 & 1.6	& 1.42$\times$10$^{22}$ & 2185$\pm$216 \\
6  & 0.8 &	 2.0	& 1.77$\times$10$^{22}$ & 2649$\pm$285 \\
7  & 0.8 & 2.25 & 2.00$\times$10$^{22}$ & 2878$\pm$58 \\
8  & 0.8 & 2.5	& 2.22$\times$10$^{22}$ & 3150$\pm$333 \\
\hline \hline
\end{tabular}
\end{center}
\tablenotetext{a}{The unit of $a_1$ is dynes $\cdot$ cm$^{-3}$ $\cdot$ Gauss$^{-1}$.}
\end{table}

\newpage
\begin{table}
\caption{\label{tab2}CME Source Region Test}
\begin{center}
\begin{tabular}{ccccc}
\hline \hline AR No.  & $\overline{B_{r}}$ along the PIL [Gauss] & $\overline{B_{r}}$ around AR Center [Gauss] & CME speed [km s$^{-1}$]\\
\hline
5   & 7.2 & 17.8 & 3096$\pm$303 \\
6   & 10.5 & 22.6 & 2545$\pm$233 \\
4  &  9.7 & 24.0 & 2480$\pm$251 \\
0  &  14.0 & 44.6 & 2185$\pm$216 \\
3  &  21.4 & 55.8 & 1550$\pm$166 \\
2  &  18.3 & 72.8 & 1450$\pm$70 \\
1  &  58.1 & 118.7 & 933$\pm$140 \\
\hline \hline
\end{tabular}
\end{center}
\end{table}

\newpage
\begin{table}
\caption{\label{tab3}EEGGL Validation for Different CRs}
\begin{center}
\begin{tabular}{ccccc}
\hline \hline CR Number & AR Number  & Input  Speed [km s$^{-1}$] & Simulated  Speed [km s$^{-1}$] & Difference [\%]\\
\hline
CR 2029  & AR 10759 & 1700 & 1866$\pm$35 & 9.8$\pm$2.1\\
CR 2106  & AR 11158 & 1250 & 1450$\pm$157  & 16.0$\pm$12.6\\
CR 2125  & AR 11520 & 1500 & 1633$\pm$64 & 8.9$\pm$4.3\\
CR 2125  & AR 11515 & 1600 & 1650$\pm$175 & 3.1$\pm$10.9\\
CR 2156  & AR 12912 & 1000 & 957$\pm$146  & -4.3$\pm$14.6\\
\hline \hline
\end{tabular}
\end{center}
\end{table}

\newpage
\begin{figure}[tbh]
\includegraphics[scale=0.8]{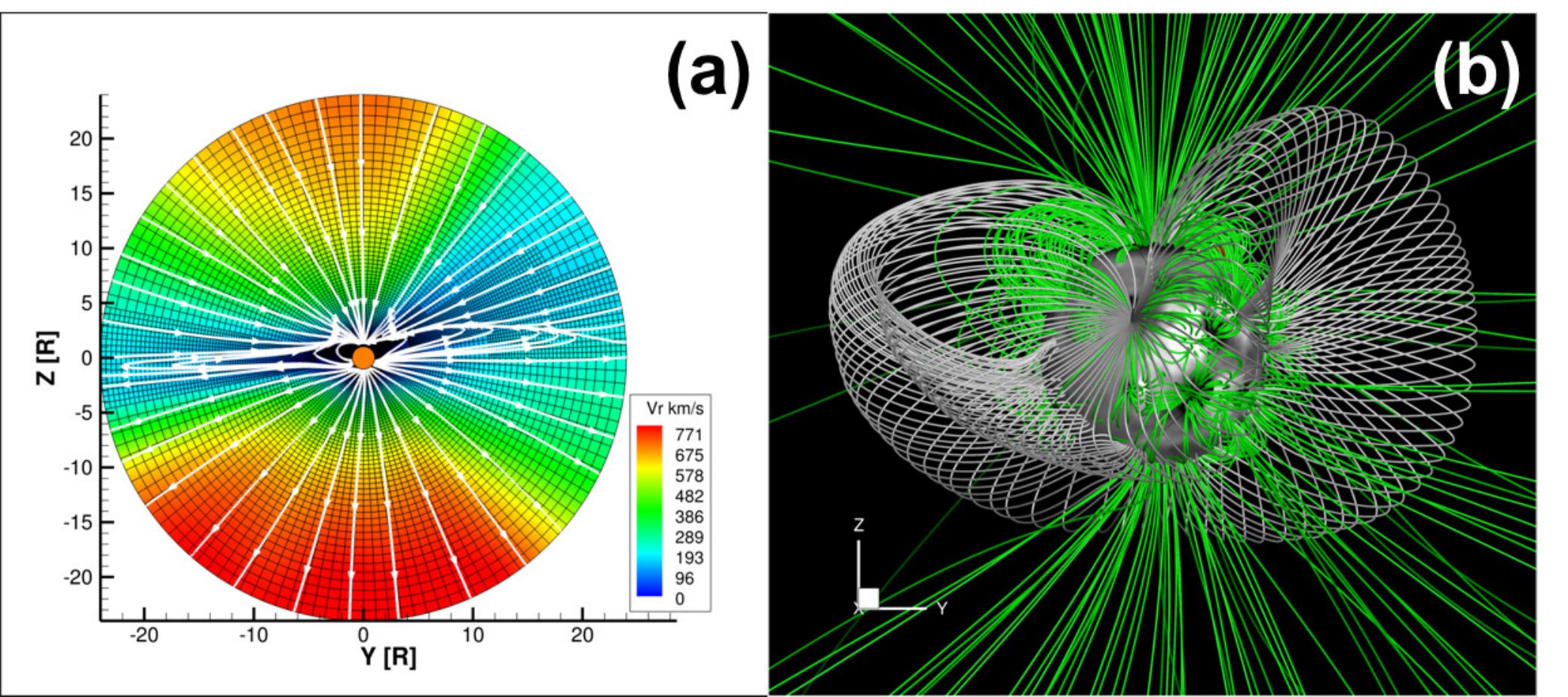}
\caption{(a) CR2107 steady state solar wind radial velocity of the meridional slice at $X=0$ with magnetic field lines. The black grid show the simulation cells. (b) 3D field configuration of the steady state solution. The white field lines represent the large-scale helmet streamer belt. The active region and open fields are marked in green.}
\end{figure}

\newpage
\begin{figure}[tbh]
\includegraphics[scale=0.85]{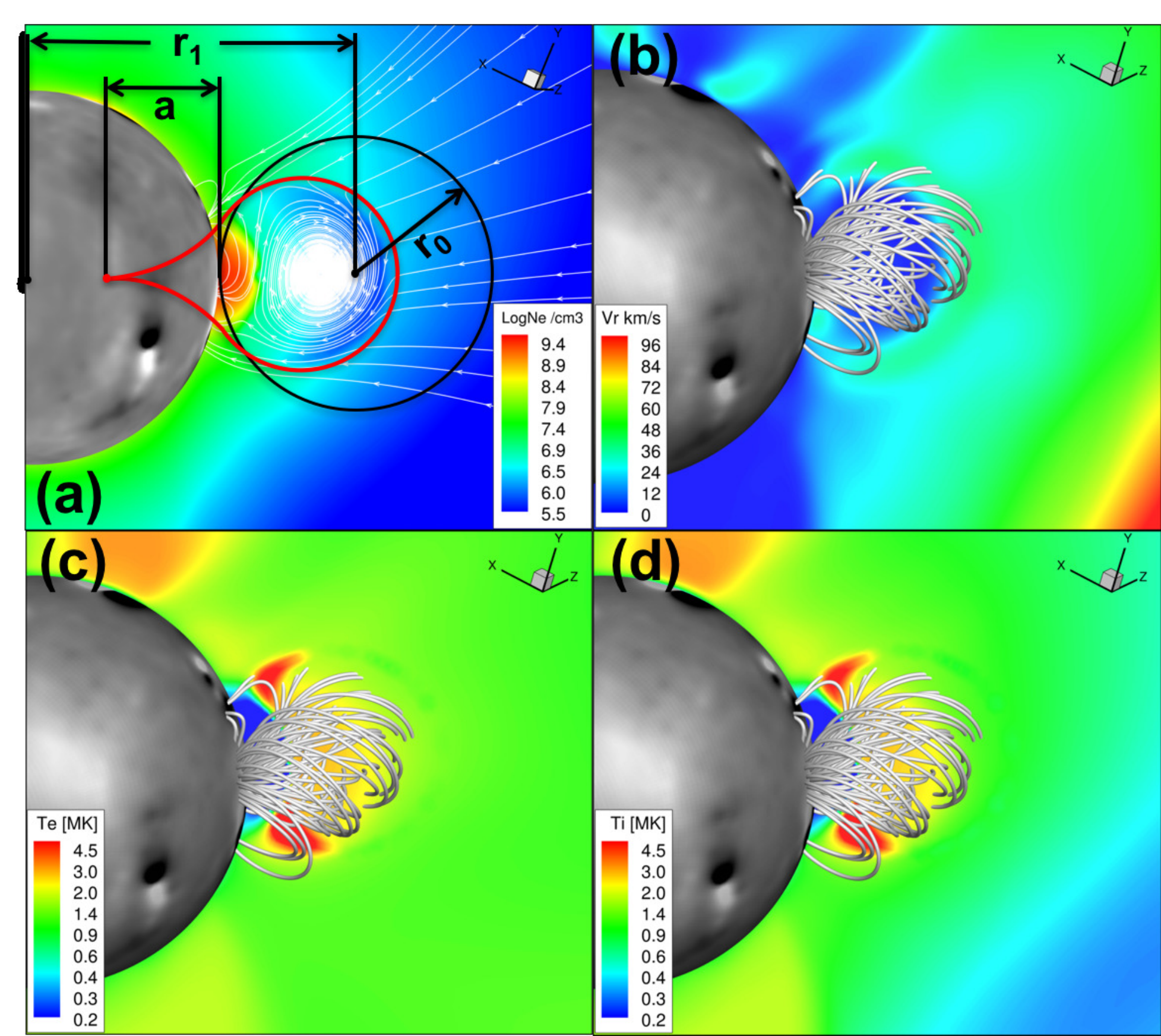}
\caption{The initial GL flux rope configuration embedded into the global solar wind solution of CR 2107. Carrington coordinates are used with Z-direction representing the rotation axis of the Sun. (a) GL flux rope in 2D with magnetic field lines and plasma density. Major GL flux rope parameters are illustrated. The black and red contours represent unstretched and stretched flux rope, respectively. (b)-(d): GL flux rope in 3D with central plane of radial velocity, electron temperature, and proton temperature, respectively.}
\end{figure}

\newpage
\begin{figure}[tbh]
\includegraphics[scale=0.75]{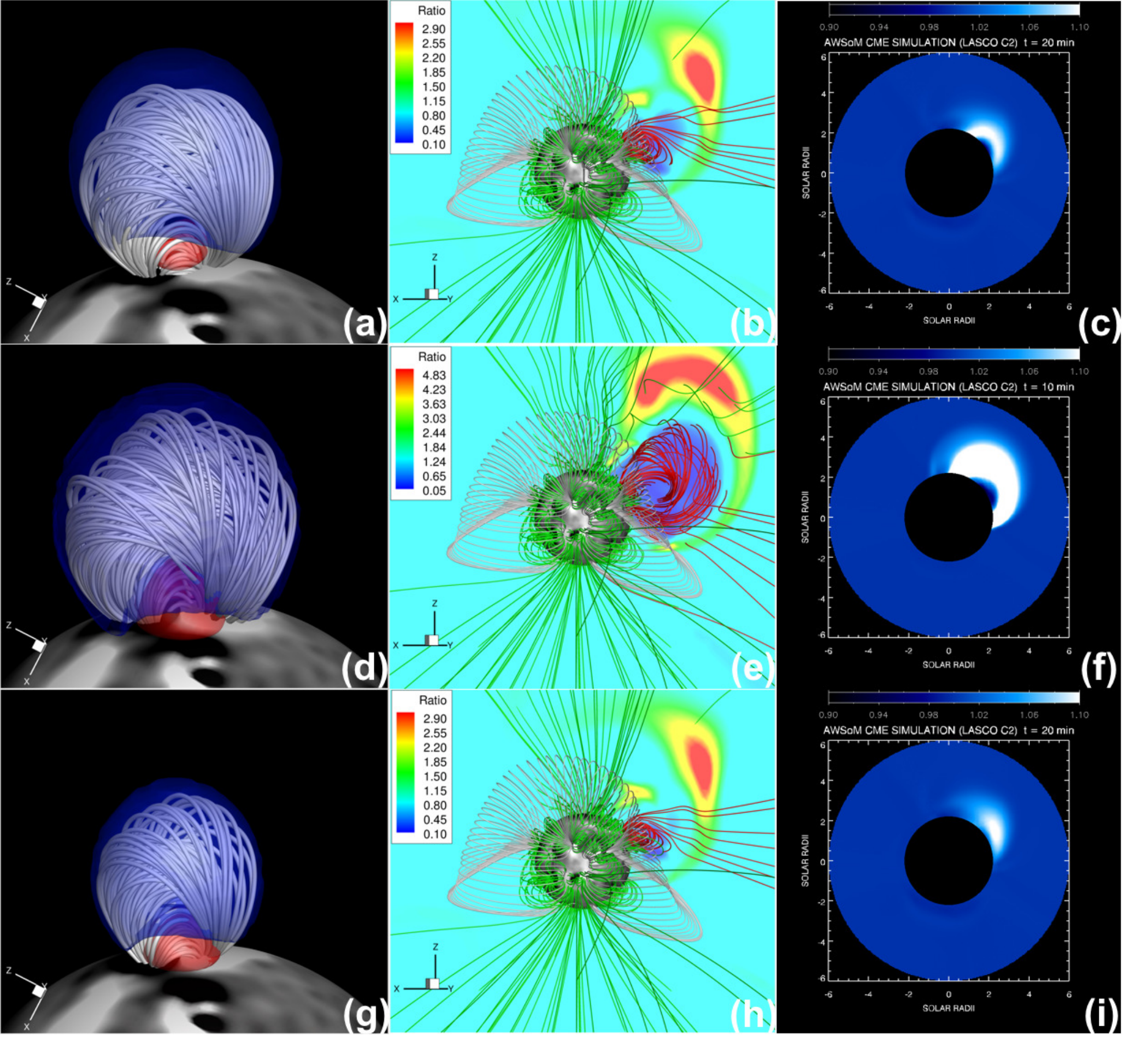}
\caption{Three examples of GL flux ropes with different size and magnetic strength parameters. (a)-(c): GL flux rope with $r_{0}=0.8$ and $a_{1}=0.6$. (a) Initial configuration of the GL flux rope. The blue and red isosurfaces represent a density ratio of 0.3 and 2.5 between the solutions after and before the GL flux rope insertion. (b) The resulting CME evolution at $t=20$ min. The background shows the density ratio between the solution at 20 min and the background solar wind. The red, white, and green field lines represent GL flux rope field lines, large-scale helmet streamers, and field lines from surrounding active regions and open field. (c) The synthesized SOHO/LASCO white light image. The color scale shows the white light total brightness divided by that of the pre-event background solar wind. (d)-(f): GL flux rope with $r_{0}=0.8$ and $a_{1}=2.25$. (g)-(i): GL flux rope with $r_{0}=0.6$ and $a_{1}=2.25$.}
\end{figure}

\newpage
\begin{figure}[tbh]
\includegraphics[scale=0.45]{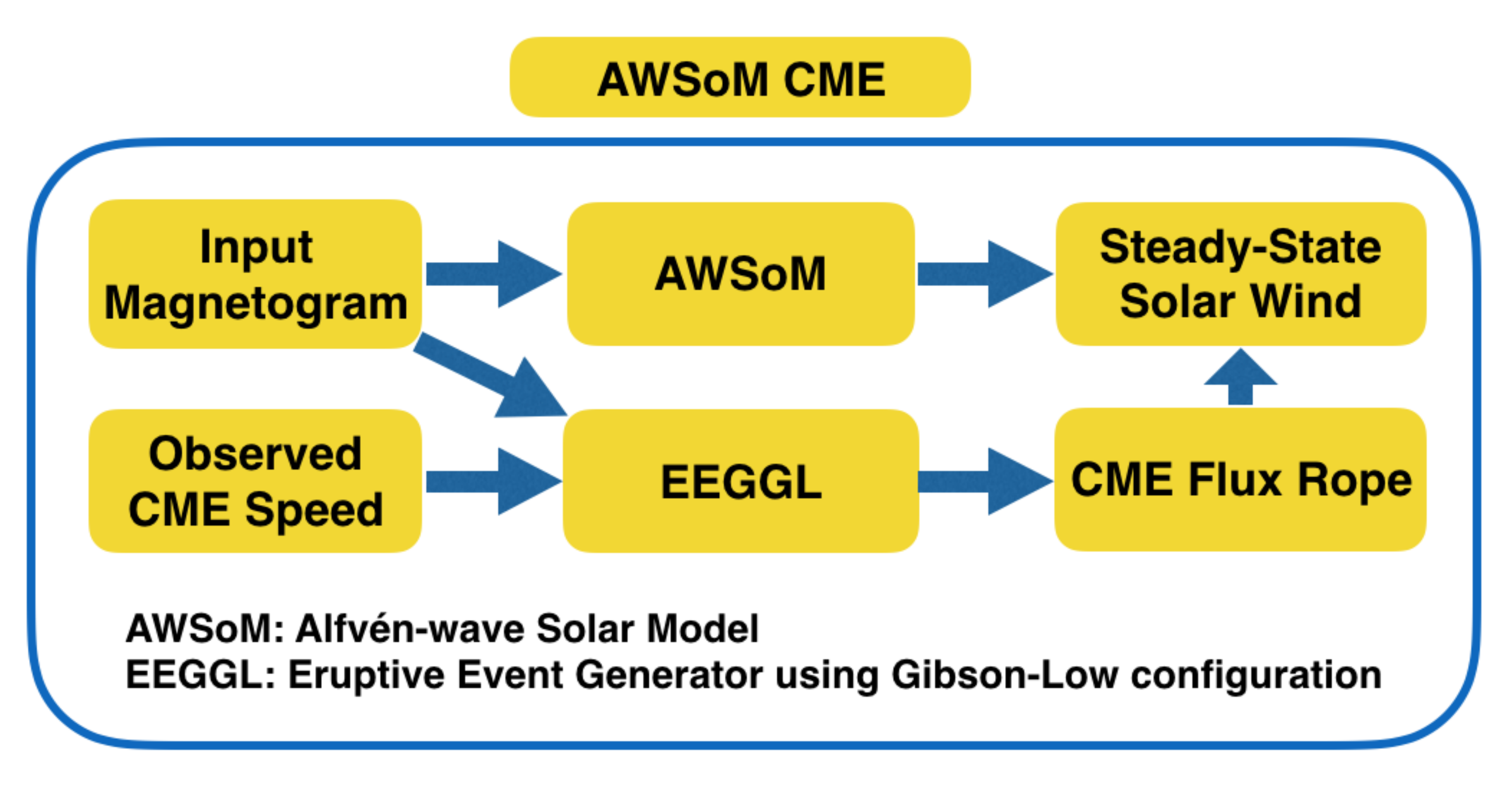}
\caption{A chart to demonstrate the procedure to simulate a CME event with AWSoM and EEGGL system.}
\end{figure}

\newpage
\begin{figure}[tbh]
\includegraphics[scale=0.65]{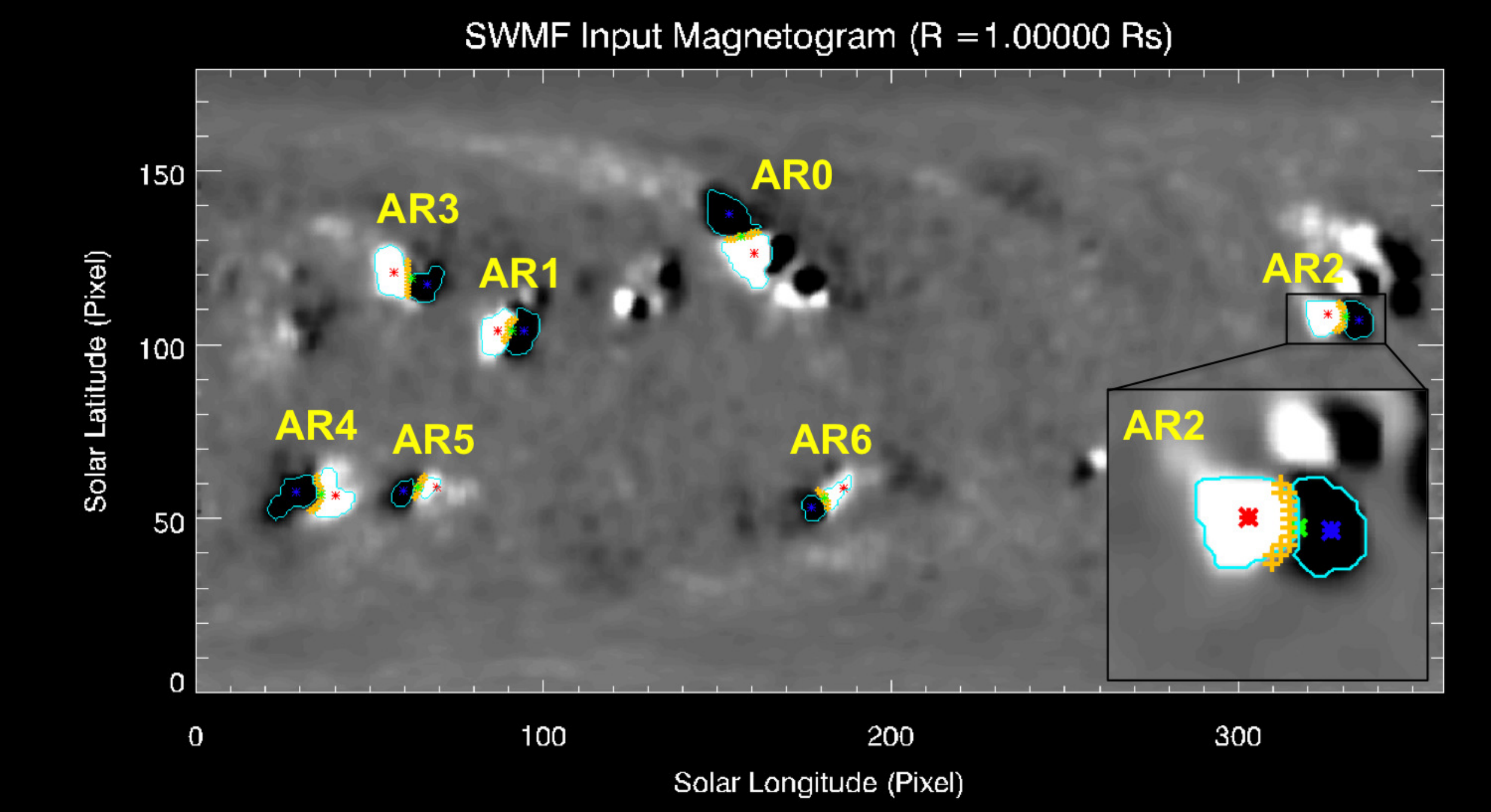}
\caption{GONG Carrington magnetogram for CR 2107. The active regions are marked with number 0 to 6. The cyan contours show the calculated active region boundaries. The red and blue symbols show the positive and negative weighted centers, respectively. The polarity inversion lines are shown in yellow and flux rope center are shown in green. A zoom-in view of AR2 is shown at the right-bottom corner of the magnetogram.}
\end{figure}

\newpage
\begin{figure}[h]
\includegraphics[scale=0.4]{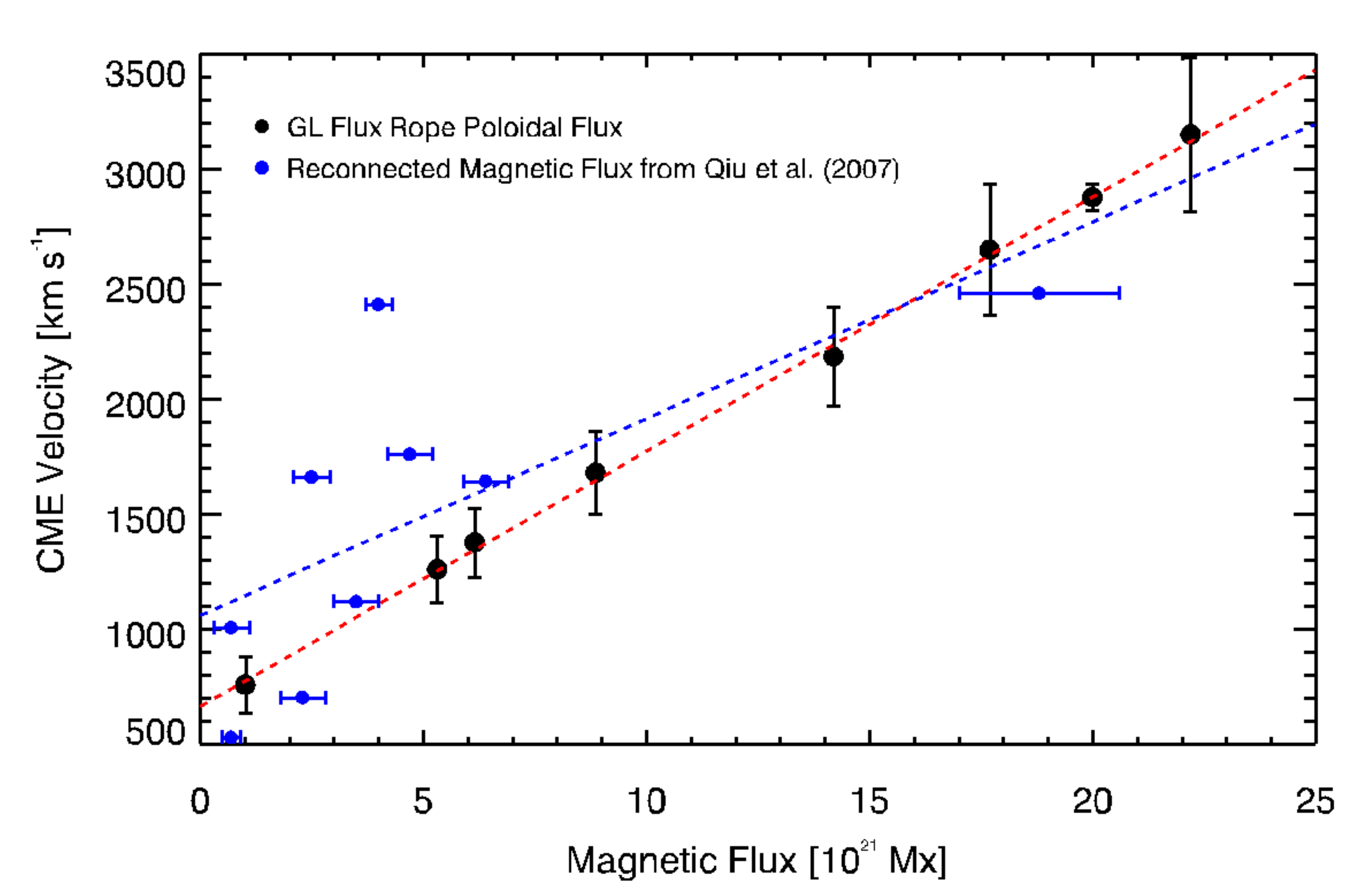}
\caption{Relationship between the poloidal flux of GL flux rope and the CME speed in the simulation (black dots). The red dashed line shows a linear fit. The corresponding data is shown in Table 1. The observational data from Qiu et al. (2007) showing the relationship between the reconnected magnetic flux and the resulting CME speed is overlaid (blue dots). A linear fit is shown in blue dashed line.}
\end{figure}

\newpage
\begin{figure}[tbh]
\begin{center}
\includegraphics[scale=0.86]{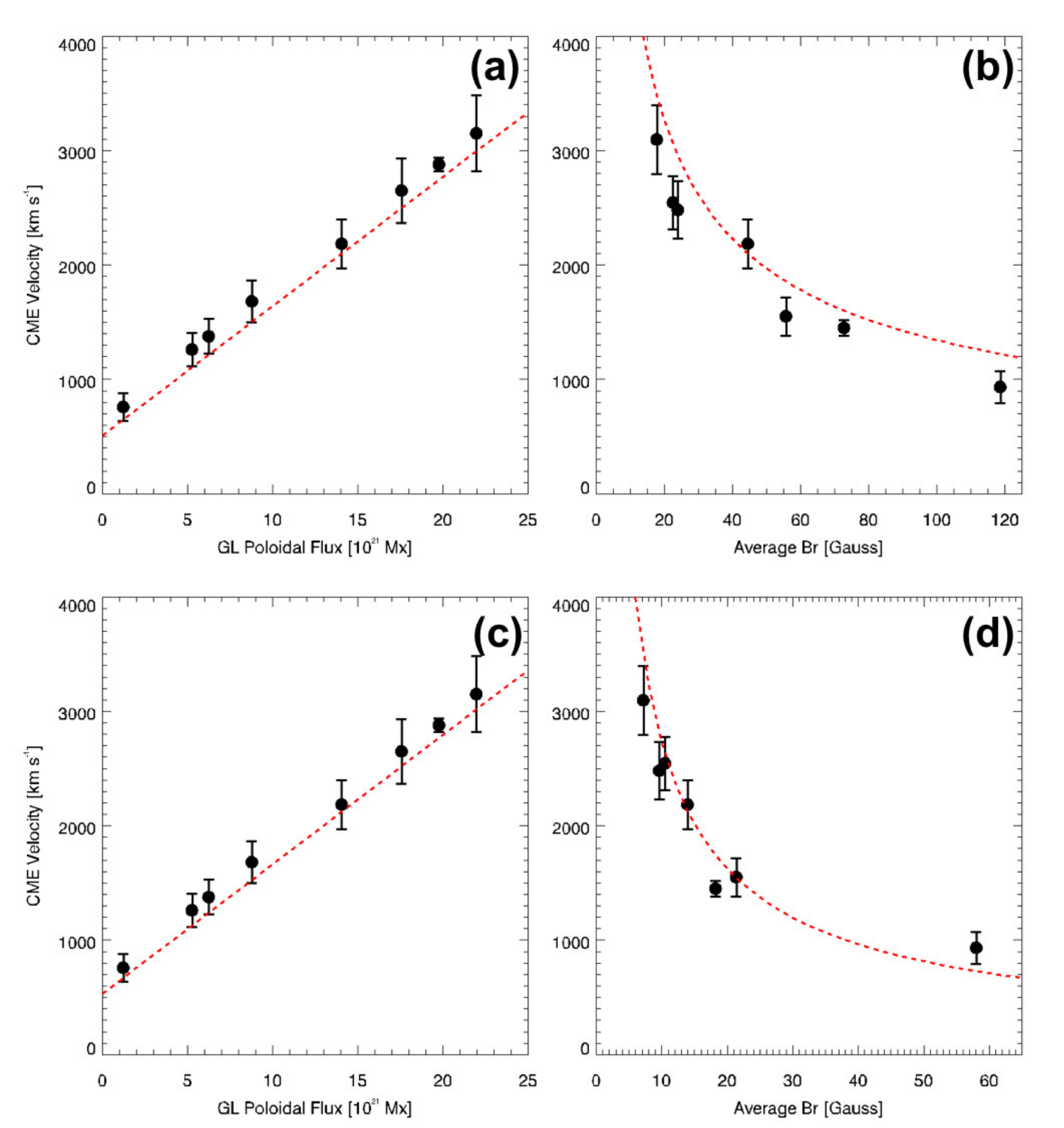}
\end{center}
\caption{Fitting curves using different definitions of active region strength. (a)-(b): Average $Br$ around the center of the active region; (c)-(d): Average $Br$ along the PIL.}

\end{figure}

\end{document}